# Surname Order and Revaccination Intentions: The Effect of Mixed-Gender Lists on Gender Differences during the COVID-19 Pandemic

Eiji Yamamura[1*], Yoshiro Tsutsui[2], Fumio Ohtake[3],

[1] Department of Economics, Seinan Gakuin University, 6-2-92 Sawaraku Nishijin Fukuoka 814-8511, Japan.
2 Faculty of Social Relations, Kyoto Bunkyo University; y.tsutsui@po.kbu.ac.jp
3 Center for Infectious Disease Education and Research, Osaka University; ohtake@cider.osaka-u.ac.jp

*Corresponding author
Email: yamaei@seinan-gu.ac.jp (EY)

## Abstract

This study probes the effects of Japan's traditional alphabetical surname-based call system on students' experiences and long-term behavior. It reveals that early listed surnames enhance cognitive and non-cognitive skill development. The adoption of mixed-gender lists since the 1980s has amplified this effect, particularly for females. Furthermore, the study uncovers a strong correlation between childhood surname order and individuals' intention for COVID-19 revaccination, while changes in adulthood surnames do not exhibit the same influence. The implications for societal behaviors and policy are substantial and wide-ranging.

# 1. Introduction

Alphabetical surname lists, often unintentionally employed by teachers in Japan owing to their convenience, can be viewed as a form of "hidden curricula" (Kawai, 1991; Matsuda, 2020; Okuno, 2022; Sudo, 2019; Takeda, 2022). Students with surnames appearing early in these lists tend to be called upon sooner in various school situations, enabling them to learn from new experiences earlier. Conversely, those with later surnames often feel marginalized as they frequently encounter new settings after their peers, impacting their future life outcomes  (Yamada, 2016). Moreover, traditionally, male students were listed before females, irrespective of alphabetical order, with same-gender students ranked by surname. This order mirrored the male-dominant societal norm, although a shift toward mixed-gender lists has occurred since the 1980s (Kawai, 1991; Matsuda, 2020; Okuno, 2022; Sudo, 2019; Takeda, 2022).

Childhood education significantly shapes precautionary and preventive behaviors against unforeseen disasters. Studies have shown that hygiene education in schools, a kind of "hidden curriculum" cultivates non-cognitive skills related to worldviews and preferences, leading to habits such as regular hand-washing and mask-wearing during pandemics like COVID-19 (Lee et al., 2022)

In the COVID-19 era, vaccinations may affect the motivation for preventive behaviors, altering individuals' actions post-vaccination  (Corea et al., 2022; Desrichard et al., 2021; Hossain et al., 2022; Rubin et al., 2021; Si et al., 2021; Wright et al., 2022; Yuan et al., 2021; Zhang et al., 2021b, 2021a). However, the motivation might differ between initial vaccination and revaccination, an area often overlooked by researchers focusing on attitudes towards vaccination (Ahmed et al., 2021; Almaghaslah et al., 2021; Brita Roy et al., 2020; Campos-Mercade et al., 2021; Carpio et al., 2021; Catma and Varol, 2021; Cerda and García, 2021; Feleszko et al., 2021; Harapan et al., 2020; Kabir et al., 2021; Lucia et al., 2021; Machingaidze and Wiysonge, 2021; Murphy et al., 2021; Qin et al., 2021; Sarasty et al., 2020; Sasaki et al., 2022; Si et al., 2021; Trogen and Caplan, 2021; Wang et al., 2021) Revaccination is crucial to sustain the protective effects of vaccination, especially as the efficacy of initial shots decreases over time.

Rate of completing the first and the second shots were approximately 78% (Digital Agency, 2023). The preventive effects of the first and second shots decreased over time. Furthermore, the third shot was effective in preventing COVID-19, especially for the newly emerged Omicron variant (Chemaitelly et al., 2021; Fleming-Dutra et al., 2022; Tartof et al., 2021; Thomas et al., 2021). Based on this evidence, the Ministry of Health, Leisure, and Welfare has publicly urged people to be revaccinated (Ministry of Health, 2022). However, the rates for third and fourth shots were approximately 68% and 46 %, respectively (Digital Agency, 2023) The rates for the third and fourth shots were distinctly lower than those for the first and second shots. Revaccination may also depend on psychological costs, even for those who complete the first and second shots. Despite the importance of revaccination, many studies have considered attitudes towards vaccination, and little is known about the issue of revaccination. This study considers how learning experiences in schools promote revaccination.

The hypothesis is that individuals with early-listed surnames may be more open to revaccination due to reduced psychological costs associated with novel situations – a benefit reaped from early school experiences. We also anticipate that the 'surname effect' would be stronger for males from the generation that did not experience mixed-gender lists, with male students invariably listed before female students. The adoption of mixed-gender lists is predicted to enhance female students' learning opportunities (Kawai, 1991; Matsuda, 2020; Okuno, 2022; Sudo, 2019; Takeda, 2022). However, to date, no quantitative research has examined the impact of adopting mixed-gender lists, an issue we address using our independently collected panel dataset.

Our findings reveal that (1) individuals with earlier childhood surnames exhibit a stronger intention to be revaccinated; (2) this correlation is larger for females if a mixed-gender list was used during their schooling; and (3) there is no correlation between adulthood surname and the intention to revaccinate. By exploring the enduring effects of implementing mixed-gender lists in schools, this study contributes to educational research, particularly in the context of unpredictable events such as the COVID-19 pandemic.

# 2. Research Background

*2.1.The alphabetical list in school*

The Japanese alphabetical order is structured around a 50-character "Kana" syllabary, classified into ten columns with five characters [1]each. In Japanese primary schools, students are listed according to their surname's alphabetical order. There are no set rules for calling on students in Japanese schools, leaving it largely up to teacher discretion. However, teachers often unconsciously rely on the alphabetical list to call on students during class and at non-academic events (Kawai, 1991; Matsuda, 2020; Okuno, 2022; Sudo, 2019; Takeda, 2022). Consequently, students with early listed surnames tend to have more frequent experiences throughout their school life. The child's surname is generally chosen from the father's name, although the mother's name can also be inherited. Regardless, students and parents cannot select an early surname from names other than their parents', indicating the surname is given exogenously.

[1] The first column consisted of 5 vowel letters, which ordered as “A,” “I,” “U,” “E,” and “O.” This is called as “A” column. Next to “A” column, “Ka” column comes, containing orderly “Ka,” “Ki,” “Ku,” “Ke” and then lastly “Ko.” In this way, 10 columns are ordered.

Japan, being a male-dominant society, has seen discriminatory practices against females in workplaces, households, and various settings. For instance, the percentage of female students attending higher-level schools was lower than that of male students, with the university entrance rate being 55.7% for males and 47.4% for females in 2017, albeit with the gender gap gradually decreasing over time (Gender Equality Bureau Cabinet Office, 2017). A possible reason for this gender gap could be disadvantages experienced by female students owing to the traditional listing method. Traditionally, male students were listed first according to their surnames, followed by female students. Therefore, all female students were listed after all male students, reflecting traditional norms in a male-dominant society.

Female students were thus less likely to develop cognitive and non-cognitive skills due to this discriminatory listing practice (Kawai, 1991; Matsuda, 2020; Okuno, 2022; Sudo, 2019; Takeda, 2022). To promote gender equality, a shift to mixed-gender lists began in the 1980s, where students were listed in alphabetical order regardless of gender. By 2017, approximately 80% of primary and junior high schools adopted this approach (Takeda, 2022).

### *2.2.Surname effects in related literatures*

Surnames are often associated with life outcomes. Alphabetical order of surnames can influence productivity and performance outcomes, with an early surname generally im-proving individual performance and evaluations. Surname order and academic performance in high schools is correlated (Autry and Barker, 1970). Additionally, surname order correlated with peer relationships among students in school life. Earlier surnames were positively correlated with various indices such as friendship, acceptance, acquaintanceship, and received liking(Marks et al., 2016). This implies that early alphabetical order contributes to forming non-cognitive skills as well as cognitive skills.

In academia, a first author with an earlier surname leads to more citations(Abramo and D'Angelo, 2017; Huang, 2015; Stevens and Duque, 2019). Earlier alphabetical order of surnames is positively correlated with better performance and evaluation, regardless of the equality of work. A similar mechanism is observed in which the paper on the cover page is more likely to be cited and highly evaluated (Battiston et al., 2019; Kong and Wang, 2020; Wang et al., 2022). A similar phenomenon is observed in the outcomes of elections. Candidates with earlier surnames have significant electoral advantages(Chen et al., 2014; Kim et al., 2015; MacInnis et al., 2021; Miller and Krosnick, 1998; Pasek et al., 2014).

One key factor is that people often choose from the top of a list, whether it be references in academic papers or a list of candidates in an election. This tendency is driven by the incentive to save time and effort when evaluating the quality of academic works or candidates' campaign pledges.

Apart from the outcome of surname advantage in school, the name is associated with firm performance (Kashmiri and Mahajan, 2014, 2010; McMillan Lequieu, 2015). For family firms, the presence of the founding family's name, as part of the firm's name, can improve performance (Kashmiri and Mahajan, 2014). In addition to performance, firms with family names have significantly higher levels of corporate citizenship and representation of customer voices (Kashmiri and Mahajan, 2010) .

There is a negative effect of surname order, which is regarded as "negative externality" in the field of economics. The inequality in labor productivity caused by surname order results in wage inequality between early and late list placements. The wage gap can potentially cause friction and conflict among workers, thus reducing incentives. A recent study found that the alphabetical order of surnames reduces team incentives, exerting a negative impact on output (Joanis and Patil, 2021).

# 3 Data and Methods

*3.1. Data collection*

Our project aimed to explore how the COVID-19 pandemic influenced individuals' preventive behaviors and subjective views on prevention at the onset of the pandemic. We aimed to achieve this by conducting repeated surveys on the same individuals. Initiated in March 2020, right after COVID-19 was detected in Japan, this project commissioned the research company INTAGE to collect data through online surveys. We selected INTAGE due to their extensive experience in academic research.

Participants were recruited by INTAGE from pre-registered individuals for the survey. The respondents were chosen randomly to meet the pre-specified quotas that represented the Japanese adult population. Data was collected on various aspects such as household income, age, gender, educational background, and area of residence. The initial survey in March 2020 saw a participation rate of 54.7%. Individuals under 18 and above 78 years were not pre-registered as subjects by INTAGE and hence were not included in the sample. Those above 78 years were considered less likely to use the internet. Therefore, the sample population was restricted to individuals between 18 and 78 years of age.

We assembled the panel data as follows: We conducted internet surveys nearly every month on 26 occasions (“waves”) between March 2020 and September 2022 with the same subjects. During this period, COVID-19 vaccines were developed and became available. These vaccines are critical in combating COVID-19, and thus are invaluable for investigating people's intentions to receive the vaccines. Many studies have explored how unvaccinated people are motivated to get inoculated against COVID-19. As of March 2023, most people have received their first and second doses of COVID-19 vaccine. It has now become crucial for vaccinated people to get revaccinated as the vaccines lose effectiveness over time. Our study focuses on revaccinations, using a sub-sample of individuals who have received their first and second doses of COVID-19 vaccines. Japanese people could receive their third dose starting from December 2021. The 19th wave of our survey was conducted the following month, in January 2022. Therefore, to examine the intention to revaccinate, we used a sub-sample covering the period from December 2021 (19th wave) to September 2022 (26th wave).

The questionnaire included queries about basic demographic, social, and economic characteristics. Descriptions of the collected variables are provided in Table 1. The outcome variable was the subjective revaccination intention, posed as the question, "Will you get a vaccination shot? Choose from 5 choices. Respondent's choices: 1(will not get a shot)-5(will get a shot)." A higher value was interpreted as indicating a higher revaccination intention. We restricted the sample to those who had received the second dose of the vaccine. Thus, in the question, we asked for their "Revaccination" intention. Even for the same individuals, their intentions varied based on the circumstances they faced. To obtain the key dependent variable, we asked in which alphabetical column their childhood (or adulthood) surnames were located. In our sample, approximately 60% of females changed their surname, compared to only 4% of males. This reflects traditional Japanese norms favoring the preservation of the male family line.

Table 1. Definition of variables used in estimations and its descriptive statistics.

| Variables | Definition | Obs. | Mean | s.d. | max | min |
|---|---|---|---|---|---|---|
| Outcome variable | | | | | | |
| *Revaccine* | Will you get a shot of vaccination? Choose from 5 choices.<br>Respondent's choices:1(will not get a shot)-5(will get a shot) | 11680 | 4.42 | 0.96 | 5 | 1 |
| Surname | | | | | | |
| *Name child* | Number of alphabetical columns of childhood surname initial ranges 1-10; 10 if respondent's childhood surname initial is in "A" column (the first column). 1 if respondent's surname initial is in "Wa" column (the 10th column). | 11680 | 7.11 | 2.45 | 10 | 1 |
| *Name child_1-5* | 1 if respondent's childhood surname initial is in "A,"",Ka", "Sa", "Ta", "Na" column (the first to fifth columns), otherwise 0. | 11680 | 0.70 | 0.45 | 1 | 0 |
| *Name Adult* | Number of columns of adulthood surname initial ranges 1-10;<br>10 if respondent's adulthood surname initial is in "A" column (the first column). 1 if respondent's surname initial is in "Wa" column (the 10th column). | 11680 | 7.10 | 2.42 | 10 | 1 |
| *Name Adult_1-5* | 1 if respondent's adulthood surname initial is in "A"",Ka", "Sa", "Ta", "Na" column (the first to fifth columns), otherwise 0. | 11680 | 0.70 | 0.46 | 1 | 0 |
| *Name adult_1* | 1 if respondent's adulthood surname initial is in "A", (the first column), otherwise 0. | 11680 | 0.22 | 0.41 | 1 | 0 |
| Primary school | | | | | | |
| *Mixed -gender list* | 1 if mixed-gender list is used in primary school, 0 otherwise.<br>Choose from 3 choices.<br>Respondent's choices:1(Yes), 2(No), 3 (Forget or will not respond). | 5248 | 0.33 | 0.46 | 1 | 0 |
| *Female teacher* | 1 if class teacher is female at in the first grade in primary school, 0 otherwise. | 11680 | 0.80 | 0.39 | 1 | 0 |
| | Respondent's choices:1(Yes), 2(No), 3 (Forget or will not respond). | | | | | |
| Control variables | | | | | | |
| *Female* | 1 if respondent is female, otherwise 0. | 11680 | 0.49 | 0.50 | 1 | 0 |
| *Ages* | Respondent's ages. | 11680 | 54.9 | 14.3 | 1 | 0 |
| *University* | 1 if respondent graduated from university, otherwise 0. | 11680 | 0.45 | 0.50 | 1 | 0 |
| *Damage* | How serious are your symptoms if you are infected with the novel coronavirus? Choose from 6 choices.<br>1 (very small influence) to 6 (death) | 11680 | 3.57 | 1.05 | 6 | 1 |
| Job status | | | | | | |
| *Office worker* | 1 if respondent is office worker, otherwise 0. | 11680 | 0.18 | 0.38 | 1 | 0 |
| *Executive* | 1 if respondent is company executives, otherwise 0. | 11680 | 0.07 | 0.25 | 1 | 0 |
| *Public officer* | 1 if respondent is public officers, otherwise 0. | 11680 | 0.05 | 0.22 | 1 | 0 |
| *Self-employment* | 1 if respondent is self-employed, otherwise 0. | 11680 | 0.05 | 0.22 | 1 | 0 |
| *Specialist* | 1 if respondent is specialist, otherwise 0. | 11680 | 0.03 | 0.17 | 1 | 0 |
| *Contract employee* | 1 if respondent is contract employee, otherwise 0. | 11680 | 0.05 | 0.22 | 1 | 0 |
| *Part-time* | 1 if respondent is part-time worker, otherwise 0. | 11680 | 0.10 | 0.32 | 1 | 0 |
| *Student* | 1 if respondent is student, otherwise 0. | 11680 | 0.01 | 0.12 | 1 | 0 |
| *Homemaker* | 1 if respondent is homemaker, otherwise 0. | 11680 | 0.22 | 0.41 | 1 | 0 |
| *No-job* | 1 if respondent does not have job, otherwise 0. | 11680 | 0.20 | 0.39 | 1 | 0 |
| *Other jobs* | 1 if respondent has other jobs, otherwise 0. | 11680 | 0.01 | 0.13 | 1 | 0 |
| Household income | | | | | 1 | 0 |
| *Income_1* | 1 if respondent's household income is below 1 million yens , otherwise 0. | 11680 | 0.02 | 0.14 | 1 | 0 |
| *Income_1.5* | 1 if respondent's household income is 1-1.99 million yens , otherwise 0. | 11680 | 0.07 | 0.25 | 1 | 0 |
| *Income_2.5* | 1 if respondent's household income is 2-2.99 million yens , otherwise 0. | 11680 | 0.13 | 0.33 | 1 | 0 |
| *Income_3.5* | 1 if respondent's household income is 3-3.99 million yens , otherwise 0. | 11680 | 0.17 | 0.37 | 1 | 0 |
| *Income_4.5* | 1 if respondent's household income is 4-4.99 million yens , otherwise 0. | 11680 | 0.14 | 0.34 | 1 | 0 |
| *Income_5.5* | 1 if respondent's household income is 5-5.99 million yens , otherwise 0. | 11680 | 0.12 | 0.32 | 1 | 0 |
| *Income_6.5* | 1 if respondent's household income is 6-6.99 million yens , otherwise 0. | 11680 | 0.09 | 0.28 | 1 | 0 |
| *Income_7.5* | 1 if respondent's household income is 7-7.99 million yens , otherwise 0. | 11680 | 0.06 | 0.24 | 1 | 0 |
| *Income_8.5* | 1 if respondent's household income is 8-8.99 million yens , otherwise 0. | 11680 | 0.04 | 0.20 | 1 | 0 |
| *Income_9.5* | 1 if respondent's household income is 9-9.99 million yens , otherwise 0. | 11680 | 0.05 | 0.22 | 1 | 0 |
| *Income_11* | 1 if respondent's household income is 10-11.99 million yens , otherwise 0. | 11680 | 0.04 | 0.19 | 1 | 0 |
| *Income _13.5* | 1 if respondent's household income is 12-14.99 million yens , otherwise 0. | 11680 | 0.03 | 0.18 | 1 | 0 |
| *Income _17.5* | 1 if respondent's household income is 15-19.99 million yens , otherwise 0. | 11680 | 0.02 | 0.15 | 1 | 0 |
| *Income_25* | 1 if respondent's household income is over 20 million yens , otherwise 0. | 11680 | 0.01 | 0.11 | 1 | 0 |
| *Time_Jan_ 2022* | 1 if the data is gathered from14 to 19 January in 2022, otherwise 0. | 11680 | 0.13 | 0.33 | 1 | 0 |
| *Time_Feb_2 022* | 1 if the data is gathered from 25 February to 2 March in 2022, otherwise 0. | 11680 | 0.13 | 0.33 | 1 | 0 |
| *Time_Apr_2 022* | 1 if the data is gathered from 15 to 20 April in 2022, otherwise 0. | 11680 | 0.13 | 0.33 | 1 | 0 |
| *Time_May _2022* | 1 if the data is gathered from 20 to 25 May in 2022, otherwise 0. | 11680 | 0.13 | 0.33 | 1 | 0 |
| *Time_June _2022* | 1 if the data is gathered from 17 to 22 June in 2022, otherwise 0. | 11680 | 0.13 | 0.33 | 1 | 0 |
| *Time_July_ 2022* | 1 if the data is gathered from 15 to 20 July in 2022, otherwise 0. | 11680 | 0.13 | 0.33 | 1 | 0 |
| *Time_Aug_ 2022* | 1 if the data is gathered from 19 to 23 August in 2022, otherwise 0. | 11680 | 0.13 | 0.33 | 1 | 0 |
| *Time_Sep_2 022* | 1 if the data is gathered from 16 to 21 September in 2022, otherwise 0. | 11680 | 0.13 | 0.33 | 1 | 0 |

Note: Numbers within parentheses are robust-standard errors.

During the sample collection process, some participants stopped taking the surveys and were dropped from the sample. We limited the samples used for analysis to respondents who participated from the 19th to the 26th waves, allowing us to follow the same individuals. In other words, all such individuals participated in the survey eight times. Table 1 shows 11,680 observations for most of the variables, indicating 1,460 unique individuals included in the sample. For a closer examination, we restricted the sample to those who answered questions about individual characteristics, including educational circumstances in primary school.

Childhood circumstances in school form non-cognitive skills that influence life outcomes (Algan et al., 2013; Ito et al., 2020; Lee et al., 2022, 2021; Yamamura, 2022a, 2022b). We purposefully collected information about the hidden curriculum. To identify the childhood surname effect, we asked about the alphabetical order of adulthood and childhood surnames. As shown in Table 1, the type of list and gender of teachers in the primacy school were included as variables to capture educational circumstances. We could only gather 5248 observations about the type of list because many respondents did not remember it. Eventually, for a closer examination, the number of respondents was reduced to 656. In Table 1, mean of *Mixed-gender list* is 0.33, indicating mixed gender list was used for 33% of respondents.

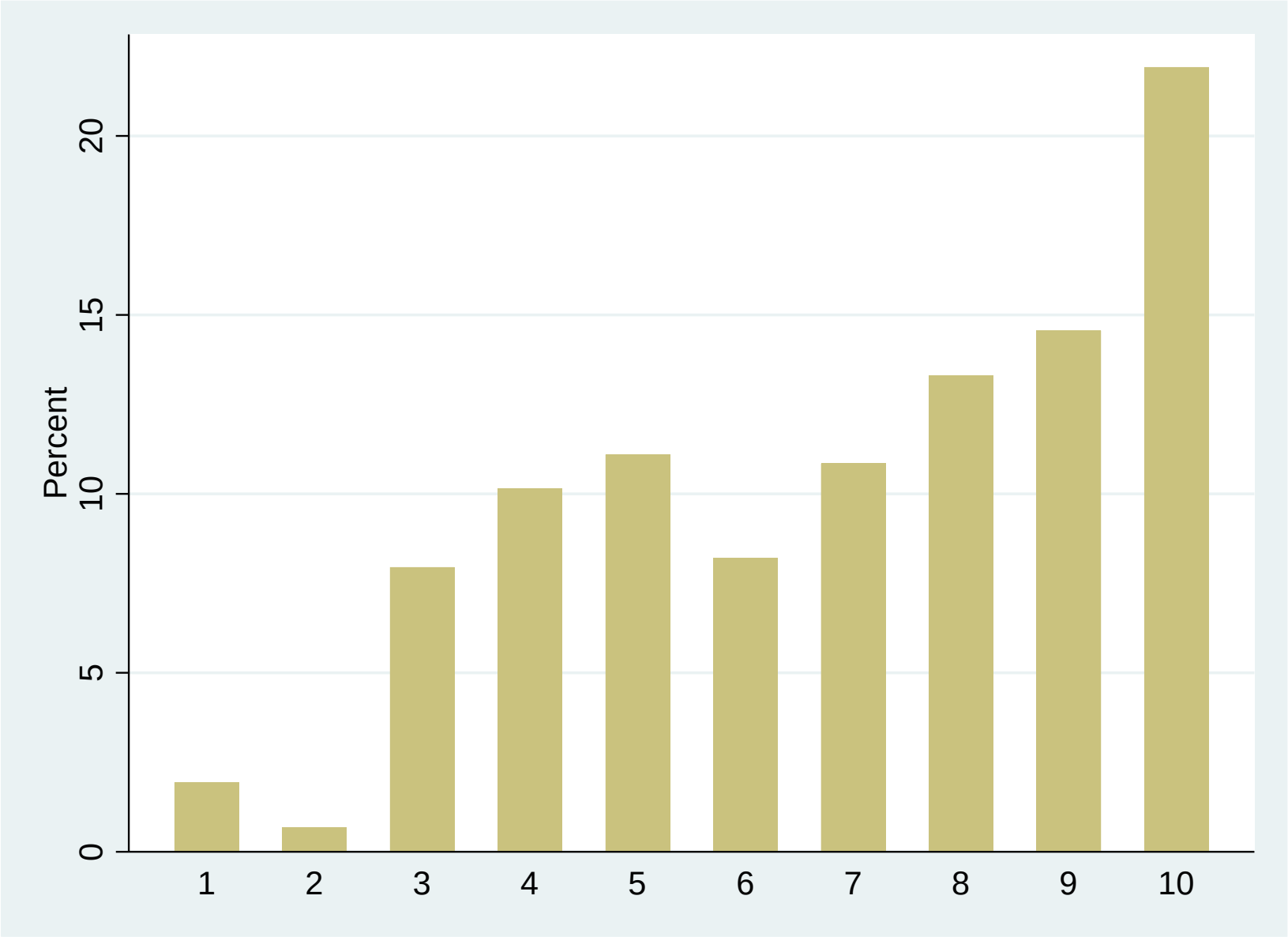

Fig 1(a) . Distribution of adulthood name order using data of general population.
Note: Data from the top 1500 family names were used to calculate percentages. This covers 9,3200,000 people, are about 75% of whole Japanese population (12,570,000).
Sources: Family Name Ranking
https://myoji-yurai.net/prefectureRanking.htm (accessed on December 18, 2022).

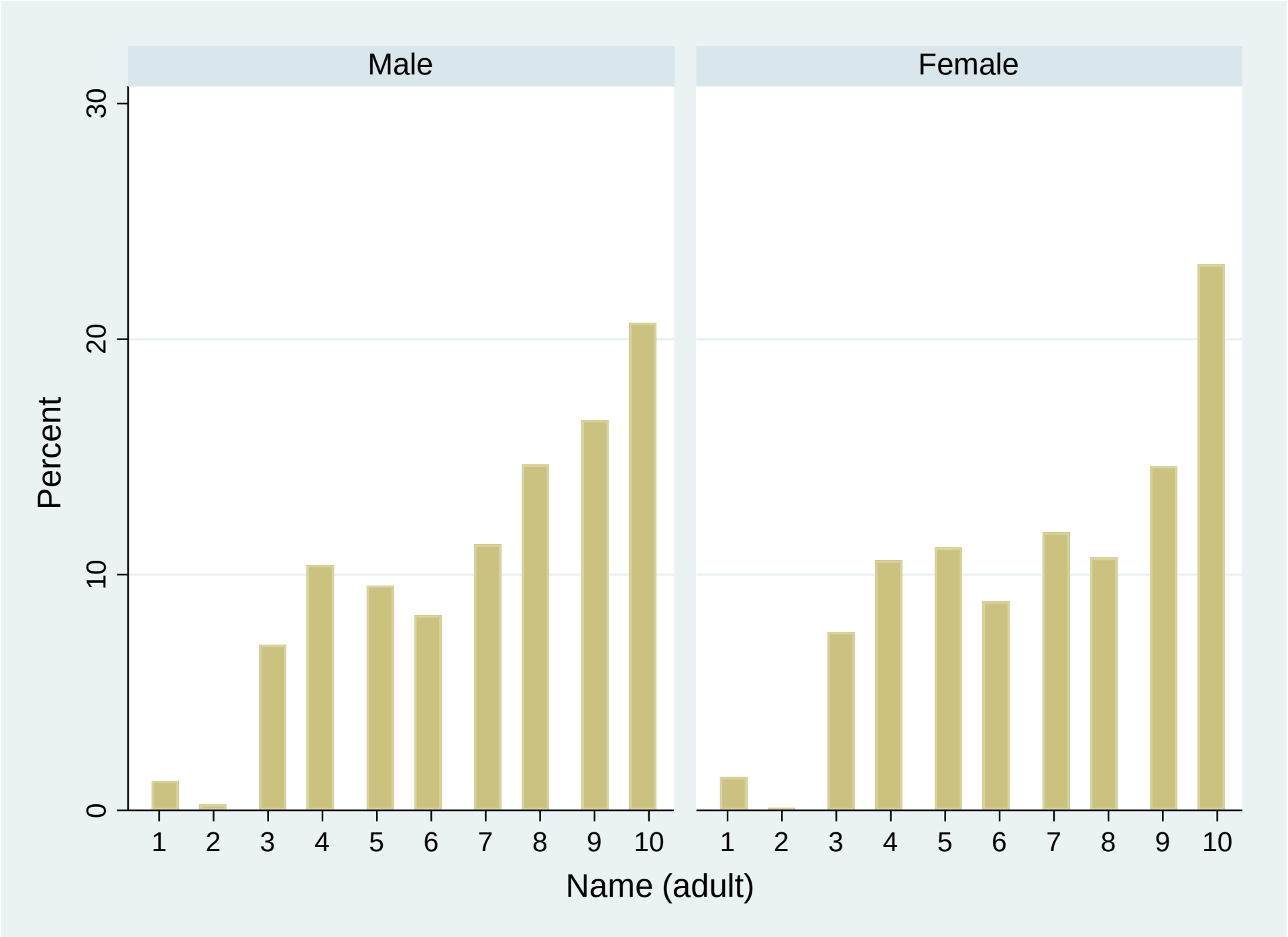

Fig 1 (b). Distribution of adulthood name order

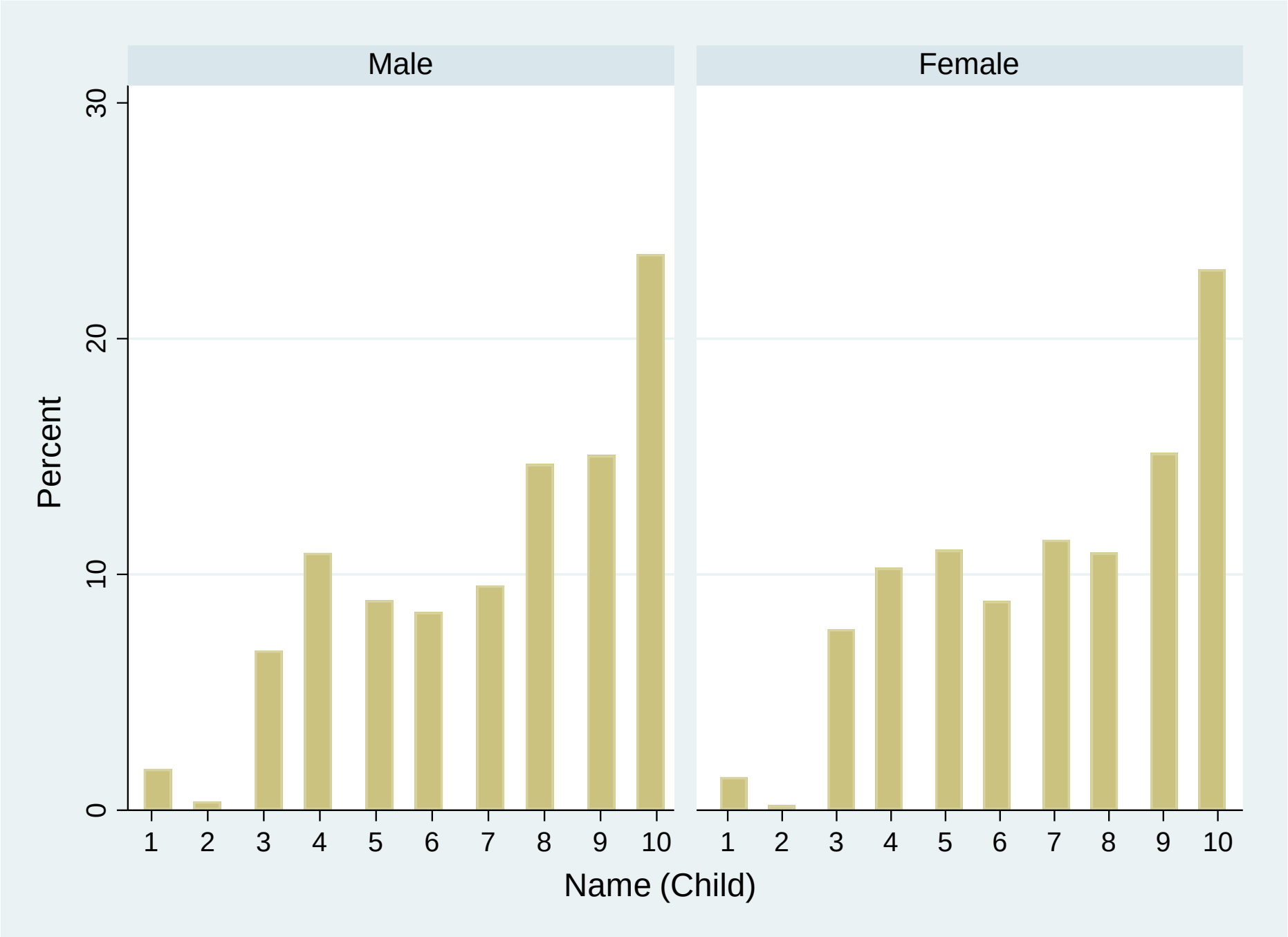

Fig 1 (c). Distribution of childhood name order

Name: For convenience of interpretation, surnames in earlier in the Japanese alphabetical list has larger values. To put it precisely, values 1, 2, 3, 4, 5, 6, 7, 8, 9 and 10 correspond to “Wa” , “Ra” , “Ya” , “Ma”, “Ha” ,”Na,” “Ta” , “Sa” , “Ka” and “A”.

The distribution of surname orders in the general Japanese population is illustrated in Fig 1 (a). For convenience of interpretation of the estimation results in a later section, the earlier the surname, the larger the number of alphabetical columns. The earliest column “A” is around 22% that is the highest ration among all columns. Using the data independently collected in this study, the distributions are shown in Figs 1 (b) and (c), which are similar to in Fig1 (a). This indicates the representativeness of the data used in this study in terms of surname distribution. Furthermore, there is no difference between males and females or between childhood and adulthood names. Hence, name distribution is not biased by gender and is influenced by life events, such as marriage. We can appropriately compare the influence of surname order between males and females, as well as between childhood and adulthood.

*3.2 Ethical issues*

The survey design in this study was conducted with ex ante approval from the Ethics Committee of the Graduate School of Economics, Osaka University. Data collection was performed in accordance with relevant guidelines and regulations. The ethics approval number of Osaka University for this study is R021014. After being informed about the purpose of the study and their right to quit at any time, the participants agreed to participate. Completion of the entire questionnaire was considered to indicate participant consent.

*3.3.Hypothesis*

It is widely acknowledged that apart from cognitive skills, non-cognitive skills generated during childhood improve various aspects of daily life(Heckman et al., 2013, 2010a, 2010b). These skills are often formed during early school experiences, helping individuals adapt to new phases in life. Even in adulthood, these non-cognitive skills help individuals cope with challenging and unfamiliar situations. We hypothesize that having an early alphabetical surname may enhance school experiences and consequently reduce the psychological cost of receiving vaccinations.

With time, vaccines tend to become less effective, even after receiving the second dose of the COVID-19 vaccination. As a result, revaccination becomes necessary to prevent the spread of COVID-19. Individuals who have already received their second shot may still choose to be revaccinated. Those whose surnames appeared early in the school register are less likely to view revaccination as troublesome. Hence, we propose Hypothesis 1: *Hypothesis 1: Having childhood earlier surname is positively correlated with the intention for revaccination.*

Compared with males, females are less inclined to be "over-confident" or to take risk (Almenberg and Dreber, 2015; Dreber et al., 2014). This is consistent with females health-related behavior in that females are more cautious and gather far more health-related information than males when they make health decisions(Ek, 2015). In fact, females were less likely than men to receive a shot of the COVID-19 vaccine than males (Daly and Robinson, 2021; Murphy et al., 2021; Ogilvie et al., 2021; Patelarou et al., 2022). There seems to be more room for female to change their attitudes toward vaccination uptake from education than for males. Therefore, learning experiences in childhood are thought to have a greater influence on females in unexpected situations such as the COVID-19 pandemic. Furthermore, adopting a mixed-gender list is expected to increase female students' opportunities because they are less likely to suffer unintended gender discrimination. Therefore, the mixed-gender list strengthened the correlation between surnames and non-cognitive skills for female students. This leads us to propose *Hypothesis 2:*

*Hypotheses 2: Females with earlier surnames are more likely to show positive attitude for revaccination if the mixed gender list was used in their school.*

As introduced in the related literature, early surnames have various advantages. However, attitudes toward revaccination are considered to depend on non-cognitive skills. Concerning non-cognitive skill formation, people with early surnames have an advantage in their childhood and school life, but not in adulthood. Using a subsample of adults whose surnames changed from childhood, we can identify the nature of surname effect. This is because the influence of their current surname is not reflected in their non-cognitive skills. Hence, we raise *Hypothesis 3.*

*Hypothesis 3: Adulthood surname is not correlated with the intention for revaccination.*

*3.4 Method*

Our model assessed how surnames relate to the outcome of the intention to revaccinate. We employed the following baseline model:

$Vaccine_{it} = \alpha_0 + \alpha_1$ *Name child_1-5 (or Name child)*$_{ih}$ $+ X_i A + e_t + u_{it}$

where $Vaccine_{it}$ is the outcome variable for individual *i* and time *h*. α denotes the regression parameters. The key variable was *Name child,* which ranged from 1 to 10. As described in Table 1, *Name child* is the number of alphabetical columns to which the respondent's childhood surname initially belonged. For convenience of interpretation, the larger the number, the earlier the surname is in the list. From *Hypothesis 1,* coefficient of *Name child* is expected to have a positive sign. However, surnames were unlikely to correlate with outcome variables. However, the correlation may not be linear. Therefore, instead of *Name child*, we included the dummy variable, *Name child 1-5* in the alternative specification. *Name child _1-5* is a dummy equal to 1 if the surname is in the first half of columns 1-5, otherwise 0.

X is a vector of various control variables. Several studies have found that the gender of teachers contributes to the formation of worldviews and preferences, which influence healthy behaviors in adulthood

(Yamamura et al.; 2019, Yamamura, 2022a; Yamamura, 2022b) . The incentive for revaccination depends on the predicted damage of COVID-19 to health if one is infected by COVID-19. Furthermore, the subjective prediction of the damage *Damage,* is likely to influence the attitude toward revaccination rather than the objective prediction. We expected the coefficient of *Damage* to have a positive value because the larger the damage, the greater the motivation for revaccination. According to a survey article on COVID-19 (Wang and Liu, 2022), the following factors are correlated with views on COVID-19 vaccination: educational background, income level, occupation, sex, and age. Hence, we also incorporated university graduation, female sex, age, household income, and job status dummies. $e_t$ captures the time-point effects when surveys are conducted. For instance, attitudes toward revaccination depend on the availability of vaccinations, the degree of spread of COVID-19, and the implemented COVID-19 policy. These effects are controlled for by including 7 time points dummies and with a default time point of January 2022. $u_{ih}$ is the error term. The ordinary least squares (OLS) model was used for estimation.

Various studies have found sex differences that correlate with precautionary behaviors. Owing to the male characteristics of overconfidence or a greater tendency to take risks (Almenberg and Dreber, 2015; Dreber et al., 2014), males may be more likely to have a positive attitude toward revaccination. During the COVID-19 pandemic, people encounter not only the risk of contracting COVID-19 but also the risk of the side effects of vaccination. In the U.S., nearly 65% of people agreed with the statement "I am worried about the side effects of the vaccine for myself," while approximately 40% people agreed with the statement "the side effects of the vaccine are likely to be worse than COVID-19 itself" (Pogue et al., 2020). People are more inclined to hesitate to receive the COVID-19 vaccination if they consider the side effects to be more severe (Brita Roy et al., 2020; Chu and Liu, 2021; Doherty et al., 2021; Kelekar et al., 2021; Kreps et al., 2021; Latkin et al., 2021; Reiter et al., 2020; Wang and Liu, 2022). The learning effects in school days possibly change the females' perception of being more positive about vaccination. To examine this, we used a specification that includes the cross-term between the female dummy and the key independent variable.

$$Revaccine_{it} = \alpha\beta_0 + \beta_1 \textit{Name child 1-5 (or Name child)}_{ih} \times \text{Female} + \beta_2 \textit{Name child 1-5 (or Name child)}_{ih} + X_i B + e_t + u_{it} \qquad \text{(model 1)}$$

The expected positive sign of coefficient of cross term, $\beta_1$, exhibits that female's name order is more strongly correlated with intention to be vaccinated in compared with male's one. We interpret $\beta_2$ as suggesting a correlation between name order and intention to vaccinate for males. Linear combination of $\beta_1 + \beta_2$ indicates the association between *Revaccine* and *Name child* for females. Furthermore, the specification, including the triple cross term, is estimated to explore how the adoption of a mixed-gender list influences the influence of name order on vaccination intention.

$$\begin{aligned} Revaccine_{it} = {} & \alpha\gamma_0 + \gamma_1 \textit{Name child_1-5 (or Name child)}_{ih} \times \textit{Female} \times \textit{Mixed-gender list} \\ & + \gamma_2 \textit{Name child_1-5 (or Name child)}_{ih} \times \textit{Female} + \\ & \gamma_3 \textit{Name child_1-5 (or Name child)}_{ih} \times \textit{Mixed-gender list} + \\ & \gamma_4 \textit{Female} \times \textit{Mixed-gender list} + \gamma_5 \textit{Name child_1-5 (or Name child)}_{ih} + \\ & \gamma_6 \textit{Female} + \gamma_7 \textit{Mixed-gender list} + \gamma_8 \textit{Name child_1-5 (or Name child)}_{ih} + \\ & X_i G + e_t + u_{it} \end{aligned}$$

Coefficient of Key variable is $\gamma_1$ shows how females learned under mixed gender list are more likely to be positive for revaccination than males who also learn under mixed gender list. In other words, it shows the degree to which the adoption of the mixed gender list reinforced the correlation between name order and revaccination intention for females compared to males. Based on *Hypothesis 2,* the expected sign of $\gamma_1$ is positive. Furthermore, even among females, the impact of name order depends on whether a mixed gender list is used. For a closer examination, the results of the linear combination are reported.

Linear combination of $\gamma_1 + \gamma_2 + \gamma_3 + \gamma_8$ indicates the correlation between *Revaccine* and *Name child* for females who learned in primary school when a gender mixed list was used. Another linear combination $\gamma_2 + \gamma_8$ exhibits the correlation for females who learned in primary school when gender mixed list was not used.

To identify the influence of name order, we used a sub-sample of those whose childhood surnames differ from those in adulthood to compare the correlation between *Name child* and *Revaccine* with that between *Name adult* and *Revaccine*. The specification of the estimated function is to add *Name adult* to the baseline model. We then simply compared the coefficient of *Name child* with that of *Name adult.* From *Hypothesis 3*, the coefficient of *Name child* is positive and its absolute value is larger than that of *Name adult*.

For the estimation, we used the statistical software Stata/MP 15.0.

# 4 Estimation Results

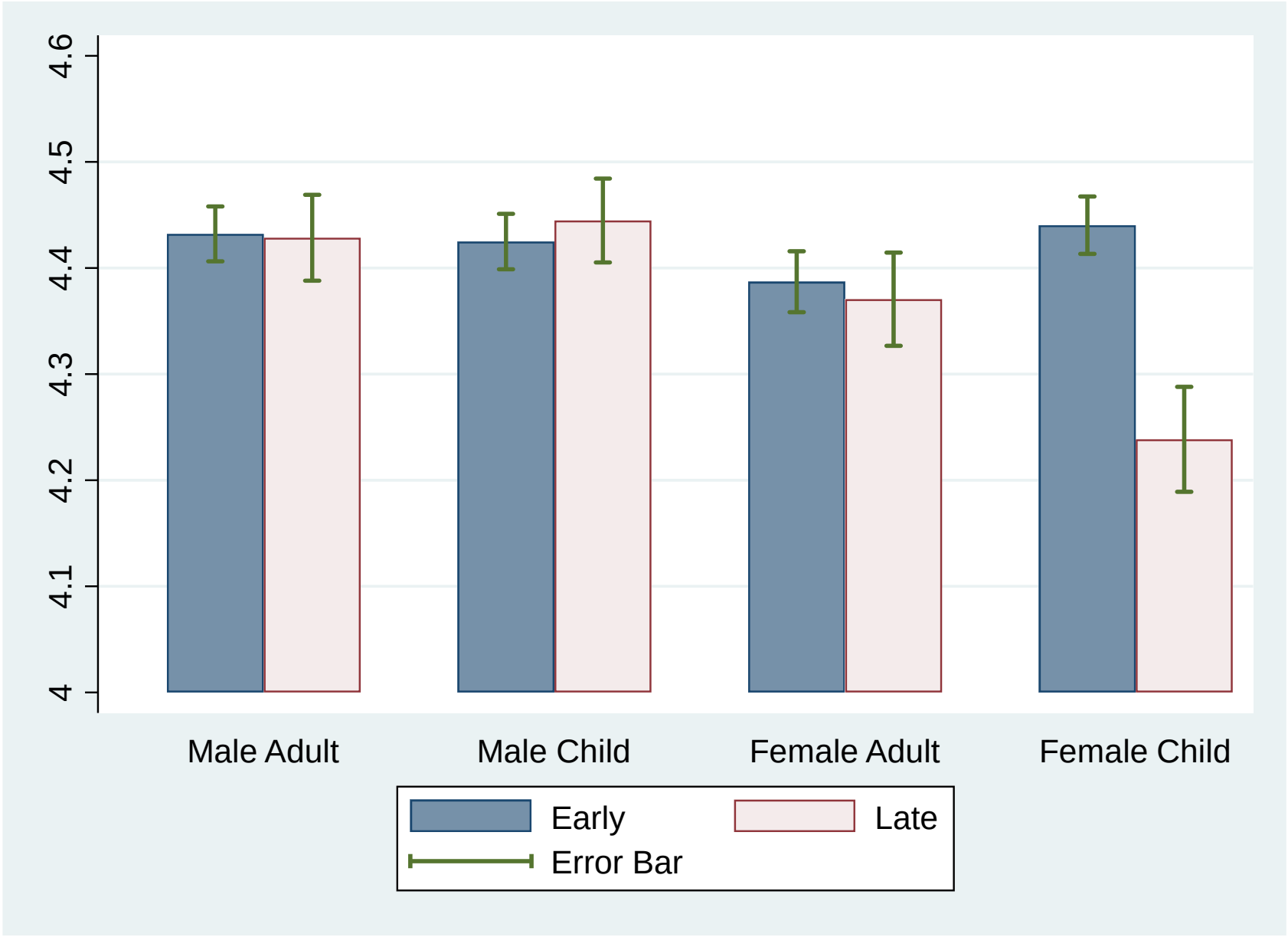

Fig. 2. Intention for revaccination between early and late surnames.
Note: There are four pairs. The sample was first divided into male and female participants. Furthermore, surnames are classified as early or later in the Japanese alphabetical list. Surname's initial in "A," "Ka," "Sa," "Ta" and "Na" is defined as "Early." Surname's initial in "Ha," "Ma," "Ya," "Ra" and "Wa" is defined as "Late."

Fig.2 demonstrates the degree of intention for revaccination by 8 groups using independently collected data. The sample was divided by gender, and either early or late childhood surname order, or adulthood name order. The surname category of individuals changed according to whether their childhood surname differed from their adult surname. The degree of revaccination intention for males was roughly higher than that for females, although this was statistically insignificant. Of particular interest is the "female-child" group on the left-hand side. Females with early childhood surnames were more likely to show higher values than those with later adulthood surnames, a statistically significant finding. Their vaccination intention was slightly higher than that of males, but the difference was not statistically significant. The findings in Fig. 2 imply that the correlation between surname order and vaccination intention was observed only for female's childhood surnames. However, various factors indicated in Table 1 were not controlled for in Fig. 2. For a closer examination, we look at the results of the regression estimations.

Table 2. Regression estimation (OLS model). Baseline model: Full sample

| | (1) | (2) |
|---|---|---|
| *Name child_1-5* | 0.095*<br>(0.042) | |
| *Name child* | | 0.022**<br>(0.008) |
| *Female teacher* | 0.043<br>(0.046) | 0.043<br>(0.046) |
| *Ages* | 0.020***<br>(0.001) | 0.020***<br>(0.001) |
| *Female* | 0.001<br>(0.044) | 0.001<br>(0.044) |
| *University* | 0.054<br>(0.035) | 0.054<br>(0.035) |
| *Damage* | 0.070***<br>(0.017) | 0.070***<br>(0.017) |
| *Office worker* | default | |
| *Executive* | 0.064<br>(0.099) | 0.064<br>(0.099) |
| *Public officer* | 0.005<br>(0.125) | 0.005<br>(0.125) |
| *Self-employment* | 0.015<br>(0.094) | 0.015<br>(0.094) |
| *Specialist* | −0.004<br>(0.149) | −0.004<br>(0.149) |
| *Contract employee* | −0.036<br>(0.103) | −0.036<br>(0.103) |
| *Part-time* | 0.036<br>(0.060) | 0.036<br>(0.060) |
| *Student* | 0.441**<br>(0.192) | 0.441**<br>(0.192) |
| *Homemaker* | −0.137<br>(0.090) | −0.137<br>(0.090) |
| *No-job* | −0.004<br>(0.074) | −0.004<br>(0.074) |
| *Other jobs* | −0.423*<br>(0.216) | −0.423*<br>(0.216) |
| *Income_1* | default | |
| *Income_1.5* | 0.087<br>(0.167) | 0.087<br>(0.167) |
| *Income_2.5* | 0.212<br>(0.181) | 0.212<br>(0.181) |
| *Income_3.5* | 0.235<br>(0.158) | 0.235<br>(0.158) |
| *Income_4.5* | 0.152<br>(0.168) | 0.152<br>(0.168) |
| *Income_5.5* | 0.162<br>(0.185) | 0.162<br>(0.185) |
| *Income_6.5* | 0.210<br>(0.195) | 0.210<br>(0.195) |
| *Income_7.5* | 0.121<br>(0.223) | 0.121<br>(0.223) |
| *Income_8.5* | 0.279<br>(0.205) | 0.279<br>(0.205) |
| *Income_9.5* | 0.285<br>(0.204) | 0.285<br>(0.204) |
| *Income_11* | 0.145<br>(0.171) | 0.145<br>(0.171) |
| *Income_13.5* | 0.443**<br>(0.171) | 0.443**<br>(0.171) |
| *Income_17.5* | 0.315<br>(0.277) | 0.315<br>(0.277) |
| *Income_25* | −0.091<br>(0.199) | −0.091<br>(0.199) |
| *Time_Jan_2022* | default | |
| *Time_Feb_2022* | −0.025<br>(0.016) | −0.025<br>(0.016) |
| *Time_Apr_2022* | −<br>0.108***<br>(0.017) | −<br>0.108***<br>(0.017) |
| *Time_May_2022* | — 0.190*** (0.023) | — 0.190*** (0.023) |
| *Time_June_2022* | — 0.275*** (0.019) | — 0.275*** (0.019) |
| *Time_July_2022* | — 0.232*** (0.020) | — 0.232*** (0.020) |
| *Time_Aug_2022* | — 0.241*** (0.020) | — 0.241*** (0.020) |
| *Time_Sep_2022* | — 0.279*** (0.023) | — 0.279*** (0.023) |
| R-square | 0.12 | 0.12 |
| Observations | 11680 | 11680 |

Note: Numbers within parentheses are robust standard errors. The Standardized EXCESS RATE is (rate of the group's population–rate of the general population)/ rate of the general population.
*p<0.10, **p<0.5, ***p<0.01

Table 2 shows baseline model using full sample, as is expected, *Name child* and *Name child_1-5* exhibit significant positive sign. Therefore, *Hypothesis 1* is supported. As for control variables, *University*, *Female*, household income dummies and job status dummies are not statistically significant, which is not congruent to previous works (Wang and Liu, 2022). In the dataset, in order to consider revaccination intention, all participants have been vaccinated. These characteristics might be correlated with whether people are vaccinated, but not with whether vaccinated people intend to be revaccinated. However, several characteristics are correlated with the intention for revaccination. Consistent with existing works (Kreps et al., 2021; Tram et al., 2022), significant positive sign of *Ages* can be interpreted as suggesting that benefit from revaccination is larger than its cost because aged people are more likely to be infected with COVID-19. In line with Kreps (Kreps et al., 2021), *Damage* shows the significant positive sign, implying that predicted seriousness of COVID-19 gives individuals incentive to revaccinate.

Table 3. Regression estimation (OLS model). Adding interaction terms between surname order and female dummy

| | (1) | (2) |
|---|---|---|
| *Name child_1-5 ×Female* | 0.205***<br>(0.005) | |
| *Name child_1-5* | − 0.024<br>(0.054) | |
| *Name child ×Female* | | 0.027**<br>(0.012) |
| *Name child* | | 0.008<br>(0.013) |
| *Female* | −<br>0.144**<br>(0.056) | − 0.195*<br>(0.098) |
| Linear combination<br>$\beta_1 + \beta_2$ | 0.181**<br>*<br>(0.050) | 0.036***<br>(0.007) |
| Control variables in Table 2 | Yes | Yes |
| R-square<br>Observations | 0.12<br>11680 | 0.12<br>11680 |

Note: Numbers within parentheses are robust standard errors. All control variables are listed in Table 2. Linear combinations are; "Name child_1-5×Female+Name_child_1_5" in column (1),
"Name child × Female + Name_child " in column (2).
*p<0.10, **p<0.5, ***p<0.01

As shown in Table 3, the significant positive sign of the cross terms implies that the correlation between surname order and revaccination intentions is stronger for females than for males. In Column (1), the absolute value of the coefficient is 0.205, suggesting that females are more likely to have a 0.205-point stronger intention on the 5-point scale than males if they have a surname in the early alphabetical order group. That is, females with early surnames had approximately 4% stronger intentions than males with early surnames.

Table 4. Regression estimation (OLS model). Adding triple interaction terms

| | (1) | (2) |
|---|---|---|
| *Name child_1-5 ×Female × Mixed-gender list* | 0.761** (0.317) | |
| *Name child_1-5 ×Female* | 0.120 (0.172) | |
| *Name child_1-5 × Mixed-gender list* | − 0.469** (0.234) | |
| *Name child_1_5* | 0.024 (0.145) | |
| *Name child ×Female × Mixed-gender list* | | 0.132** (0.063) |
| *Name child ×Female* | | 0.027 (0.037) |
| *Name child × Mixed-gender list* | | − 0.063 (0.054) |
| *Female × Mixed-gender list* | − 0.628*** (0.209) | − 1.003** (0.434) |
| *Name child* | | 0.008 (0.033) |
| *Female* | − 0.044 (0.121) | − 0.165 (0.257) |
| *Mixed-gender list* | 0.235 (0.167) | 0.356 (0.379) |
| Linear combination I $\gamma_1+\gamma_2+\gamma_3+\gamma_8$ | 0.436** (0.207) | 0.103*** (0.035) |
| Linear combination II $\gamma_2+\gamma_8$ | 0.144 (0.105) | 0.034** (0.016) |
| Control variables in Table 2 | Yes | Yes |
| R-square | 0.16 | 0.16 |
| Observations | 5248 | 5248 |

Note: Numbers within parentheses are robust standard errors. All control variables are listed in Table 2.
Linear combination I is; "Name child_1-5×Female×Mixed-gender list+ Name child_1-5×Female+ Name child_1-5×Mixed-gender list +Name_child_1_5" in column (1), "Name child×Female × Mixed-gender list+ Name child× Female+ Name child×Mixed-gender list + Name_child" in column (2).
Linear combination II is; "Name child_1-5×Female +Name_child_1_5" in column (1), "Name child× Female + Name_child" in column (2).
Linear combination I shows the name order effect on female learning in primary school when a mixed-gender list is adopted. Meanwhile, Linear combination II shows the name order effect for female students in primary school when a mixed gender list was not adopted.
*p<0.10, **p<0.5, ***p<0.01

Table 4 shows results of triple cross term to examine the impact of mixed gender list.

*Name child_1-5×Female×Mixed-gender list* indicates a positive sign and statistically significant 5 % level in Column (1). The cross term in Column (2) also shows the significant positive sign. Correlation between early surname and revaccination intention is larger for females when mixed gender list was used in their school, than when mixed gender list was not used. Further, results of "Linear combination I" in Column (1) indicate that females in early surname group have 0.436-point stronger intention than males if the mixed gender list was used. Further the difference is statistically significant. However, we see from "Linear combination II" that the statistical difference disappeared if the mixed gender list was not used. In alternative specification in Column (2), regardless of mixed gender list, female's early name order is more strongly correlated with revaccination intention than males. However, absolute value of its coefficient is about three times larger if mixed gender list was used than if mixed gender list was not used. Considering results jointly leads us to argue that adoption of mixed gender list strengthened early name influence for females. This is consistent with the *Hypothesis 2.*

Table 5. Regression estimation (OLS model).

Comparing childhood and adulthood surname. Sub-sample of childhood surname differing from adulthood one.

| | Male & Female | | Female | |
|---|---|---|---|---|
| | (1) | (2) | (3) | (4) |
| *Name child_1-5* | 0.175**<br>(0.073) | | 0.198**<br>(0.073) | |
| *Name adult_1_5* | −0.063<br>(0.072) | | −0.102<br>(0.076) | |
| *Name child* | | 0.039***<br>(0.011) | | 0.045***<br>(0.011) |
| *Name adult* | | −0.001<br>(0.013) | | −0.009<br>(0.014) |

| | | | | | |
|---|---|---|---|---|---|
| Control variables in Table 2 | Yes | Yes | | Yes | Yes |
| R-square | 0.14 | 0.14 | | 0.15 | 0.15 |
| Observations | 3640 | 3640 | | 3408 | 3408 |

Note: Numbers within parentheses are robust standard errors. All control variables are listed in Table 2.
*p<0.10, **p<0.5, ***p<0.01

To compare childhood surname impact with adulthood one, *Name adult_1_5* and *Name adult* were included to baseline specification. Further, the estimation was conducted using a subsample of those whose childhood surnames are different from those in adulthood. After marrying, women generally change their surnames, whereas male do not. Thus, males in the subsample can be considered different from the general male population. Hence, for the estimation, we also used a subsample excluding males, and the results are reported in Columns (3) and (4). For *Name adult_1_5* and *Name adult,* we did not observe a positive sign or statistical significance in any column. Meanwhile, *Name adult_1_5* and *Name adult* indicate significant positive signs in all columns. These results imply that the surname order in childhood leads to differences in revaccination intentions. This finding strongly supports *Hypothesis 3*.

# 5 Discussion

### *5.1 Implication*

The purpose of this study was to examine how surname order forms non-cognitive skills that influence gender differences in revaccination intentions. Additionally, we examined the effect of the education policy of adopting a mixed gender list on long-term health behaviors such as revaccination.

The estimation results can be interpreted from the viewpoint of a cost-benefit analysis.

Guaranteed payments for COVID-19 vaccination have increased vaccination rates (Campos-Mercade et al., 2021)(Campos-Mercade et al., 2021)(Campos-Mercade et al., 2021). Therefore, people are hesitant about vaccination if the cost of vaccination is greater than its benefit. In other words, the benefit outweighs the cost for those who receive a vaccination shot. The vaccinated individuals agreed with the benefit of vaccination when they received at least the first and second shots. However, their attitudes towards revaccination varied. One reason for this is that the subjective and psychological costs of revaccination differ. The key findings of this study show that abundant experience under new phases in school days reduces the likelihood of repeating the same things, such as revaccination.

Many studies have found that females are hesitant to COVID-19 vaccination(Daly and Robinson, 2021; Murphy et al., 2021; Ogilvie et al., 2021; Patelarou et al., 2022). Lee et al. (Lee et al., 2022) found that informal school education contributed to the promotion of preventive behaviors against COVID-19. However, it is unknown how the gender gap in vaccination intentions can be reduced. This study provided evidence that insufficient experience during new phases in school causes females' reluctance to revaccinate.

*5.2 Strength*

Various studies have considered the influence of the name order. However, this study is the first to analyze the outcome of name order by bridging education economics and public health. The name order list was used as part of the hidden curriculum, which resulted in gender discrimination. The mixed-gender list has been gradually adopted since the 1980. This study shows that the mixed gender list in schools reduced the gender gap in adulthood by considering revaccination intentions. A mixed gender list is useful for forming non-cognitive skills among females, and its influence persists even when they become adults.

*5.3 Limitation*

We assumed that the surname is given exogenously; thus, the setting is a natural experiment. However, the effect of the surname may have been inherited from one generation to another. The distribution of surnames among the elite and underclass indicates that social status is inherited(Clark and Cummins, 2014). Surnames reflecting inherited physical aptitude are advantageous for certain sports activities (Bäumler, 1980; Voracek et al., 2015). Owing to data limitations, it was difficult to distinguish inherited effects from learning experiences in schools. However, physical characteristics and social status seem to be irrelevant to revaccination intention. Furthermore, worldwide pandemics such as COVID-19 have not occurred since the Spanish flu around 100 years ago. Therefore, the current generation is unlikely to be heavily influenced by parental experience.

We assumed that the adoption of a mixed-gender list is determined exogenously. However, the adoption of such a list could be related to teacher quality. If this holds true, the causality is unclear. Consequently, we used the mixed-gender list as a proxy for teacher quality. Female teachers are less likely to discriminate against students of the same gender. We controlled for the gender of the homeroom teacher. To a certain extent, teacher quality is thus controlled, although we aim to control for teacher quality more precisely in future studies. Recall bias may occur when considering the type of list used in primary school, but in the questionnaire, respondents could choose "Forget or will not respond," which mitigates this bias.

Although the additional contribution is minor for examining general COVID-19 vaccination intention, as numerous studies have already scrutinized it, we confined our sample to those who completed the first and second shots of the COVID-19 vaccination, aiming to consider intentional revaccination. Naturally, the estimation results may suffer from selection biases. However, as of March 2023, the vaccination rate was approximately 80%, implying that the effect of selection bias is likely small (Digital Agency, 2023). Considering the policy implications at the stage when vaccinated people are the majority, using a subsample is sufficiently valuable.

# 6 Conclusion

Surnames are given exogenously and collecting surname data allows researchers to conduct quasi-experiments. The surname effect has been considered in various fields including academic performance, election outcomes, and firm performance. The contributions of this study are as follows.

First, no study has analyzed the connection between the role of surnames in schools and educational outcomes in later life. This study is the first to compare childhood and adulthood surname orders to identify how surname order plays a critical role in forming non-cognitive skills.

Second, educational researchers and mass media have asserted that mixed gender lists should be adopted to equalize learning opportunities between genders in school life(Kawai, 1991; Matsuda, 2020; Okuno, 2022; Sudo, 2019; Takeda, 2022). In response to this claim, the rate of adoption mixed gender lists has increased to 80 % in primary and junior high schools. However, no studies have quantitatively verified the influence of adopting a mixed-gender list. We contribute to education research by showing the long-term effect of adopting a mixed-gender list in schools when unexpected events such as the COVID-19 pandemic occur.

The implication of the major findings is that gender differences are partly owing to traditional name order lists in schools. This list leads teachers to unintentionally discriminate against female students by reducing their learning experiences. Insufficient female experience in school increases the psychological cost of revaccination, lowers the benefit of revaccination, and delays the eradication of COVID-19.

While we provided evidence that a mixed-gender list is effective in reducing the gender gap in school experiences, this study deals with a very specific issue and derives a general argument. Thus, it's necessary to explore the impact of mixed-gender lists on broader social and economic issues. These remaining concerns should be addressed in future studies. Further examination of how childhood surname order influences the quality of adult life is warranted.

# Acknowledgments

We thank Editage (http://www.editage.com) for editing and reviewing the manuscript for English language.